\documentclass{fundam}

\usepackage{amsfonts,amssymb,amsmath}
\usepackage{wasysym}
\usepackage[dvips]{graphics}
\usepackage{graphicx}
\usepackage{clrscode}
\usepackage{tikz}
\usepackage{epsfig}

\usepackage{url}
\usepackage{latexsym}

\definecolor{aocolour}{rgb}{0.7,0.8,1}
\definecolor{mkcolour}{rgb}{1,0.9,0.7}

\if01
\newtheorem{theorem}{T\/heorem}[section]
\newtheorem{apptheo}{T\/heorem}[section]
\newtheorem{corollary}{Corollary}[section]
\newtheorem{definition}{Definition}[section]
\newtheorem{lemma}{Lemma}[section]
\newtheorem{applem}{Lemma}[section]
\newtheorem{example}{Example}[section]
\newtheorem{comment}{Comment}[section]
\newtheorem{fact}{Fact}[section]
\newtheorem{claim}{Claim}[section]
\newtheorem{proposition}{Proposition}[section]
\newtheorem{remark}{Remark}[section]

\newtheorem{open}{Open problem}[section]
\newtheorem{theorem}{T\/heorem}

\newtheorem{definition}[theorem]{Definition}

\fi

\usepackage{hyperref}

\begin{document}


\setcounter{page}{231}
\publyear{24}
\papernumber{2180}
\volume{191}
\issue{3-4}

\finalVersionForARXIV


\title{Descriptional Complexity of Finite Automata -- Selected Highlights }

\author{Arto Salomaa\\
 Department of Mathematics and Statistics\\
 University of Turku, 20014 Turku, Finland\\
	asalomaa@utu.fi
\and
Kai Salomaa\thanks{Address for correspondence:  School of Computing, Queen's University, Kingston, Ontario, Canada.}
   \\
School of Computing\\
 Queen's University,  Kingston, Ontario, Canada\\
salomaa@queensu.ca
\and
Taylor J. Smith \\
Department of Computer Science\\
 St. Francis Xavier University, Antigonish, Nova Scotia, Canada\\
tjsmith@stfx.ca
}

\maketitle

\runninghead{A. Salomaa et al.}{Descriptional Complexity -- Highlights}

\vspace{-8mm}
\begin{flushright}
	{\em In honor of the 60th birthday of Iiro Honkala.\hspace*{10mm}}
\end{flushright}\vspace*{-5mm}

\begin{abstract}
	The state complexity, respectively,
	nondeterministic state complexity of a regular language $L$ is the
	number of states of the minimal deterministic, respectively,
	of a minimal nondeterministic finite automaton for $L$.
	Some of the most studied state complexity questions deal
	with size comparisons of nondeterministic finite automata
	of differing degree of ambiguity. More generally, if
	for a regular language
	we compare
	the size of description  by a finite
	automaton and by a more powerful language definition mechanism,
	such as a context-free grammar, we encounter
	 non-recursive
	trade-offs.
	Operational state complexity studies the state complexity of
	the language resulting from a regularity preserving operation
	as a function of the complexity of the argument languages.
	Determining the state complexity of combined operations is
	generally challenging and for general combinations of operations
	that include intersection and marked concatenation it is
	uncomputable.

\end{abstract}

\section{Introduction}

Descriptional complexity studies
the relative succinctness between different representations
of formal languages \cite{Hartmanis1983}.
For a quantitative understanding of regular languages the
commonly used size measures count the
number of states or, in the case of nondeterministic
finite automata, the number of transitions.
Early work
on  descriptional complexity of finite automata
includes \cite{Lupanov1963,Maslov1970,Meyer1971,Moore1971,Stearns1967}.

\medskip
One of the most important open problems in descriptional complexity
was originally raised by Sakoda and Sipser in 1978
\cite{Sakoda1978}.
The question asks whether any two-way nondeterministic finite automaton $M$ has
an equivalent deterministic two-way automaton with a number of
 states bounded by a polynomial on the number of states of $M$.
Already Sakoda and Sipser conjectured a negative answer to this question.
Berman \cite{Berman1980} and Sipser \cite{Sipser1980} showed that
if one proves an exponential gap in the succinctness of nondeterministic
and deterministic two-way automata and the strings involved in the
separation have polynomial length, this implies that deterministic
logarithmic space is a proper subset of nondeterministic logarithmic space.
Also Sipser \cite{Sipser1980} introduced
a restricted version of two-way automata, called sweeping automata,
where the reading head may reverse only at the end-markers and proved
an exponential separation between nondeterministic and deterministic
sweeping automata.
While the original question on the succinctness comparison of nondeterministic
and deterministic two-way automata remains open, an exponential separation
has been established, besides sweeping automata, for several
other restricted variants \cite{Hromkovic2003,Kapoutsis2014,Pighizzini2013}.

This brief survey  discusses three specific descriptional
complexity topics:
non-recursive trade-offs, the state complexity trade-off between
nondeterministic finite
automata  with differing
degrees of ambiguity, and, the state complexity of
language operations. We generally focus on finite automaton models,
however, non-recursive trade-offs naturally deal with succinctness
comparisons with more powerful models.
Descriptional complexity is a large and active research area and more
information can be found e.g. in the surveys
\cite{Gao2017,Goldstine2002,Gruber2015,Gruber2021,Holzer2011b,Hromkovic2002b,Pighizzini2013,Yu1997}.

\section{Non-recursive trade-offs}

In a seminal work
Stearns \cite{Stearns1967} studied the relative succinctness of
regular languages represented by deterministic finite automata (DFA)
and deterministic pushdown automata (PDA). He showed that if a deterministic
PDA recognizes a regular language it can be a simulated
by a DFA of triple-exponential size. The work  establishes also the
decidability of regularity of the language of a deterministic
PDA. Generally the difference in succinctness of description
between different representations of a regular
language can be arbitrary. This phenomenon is
referred to as a non-recursive trade-off.

\medskip
More formally, a family of languages ${\cal L}$ is represented by a
descriptional system $S$ if ${\cal L} = \{ L({\cal R}) \mid  {\cal R}
\in S \}$. Here $L({\cal R})$ is the language represented by
descriptor ${\cal R}$.
A complexity measure is a total recursive function
$c : S \rightarrow \mathbb{N}$.
For descriptional systems $S_1$ and $S_2$ equipped with a complexity
measure $c$, a function $f$ is said to be an upper bound for the
size blow-up for changing from system $S_1$ to $S_2$ if for
every language $L$ that has representations
in both systems, for every representation ${\cal R}_1$ of $L$ in $S_1$,
the language $L$ has a representation ${\cal R}_2$ in $S_2$ where
$c({\cal R}_2) \leq f(c({\cal R}_1))$. Note that, for example, when
considering the
trade-off between finite automata and pushdown automata we consider
only the size of representations of regular languages. The trade-off
between two descriptional systems is said to be {\em non-recursive\/}
if it is not upper bounded by any recursive function.

\medskip
A central part of descriptional complexity research
deals with  non-recursive trade-offs.
As a first non-recursive
size blow-up, Meyer and Fischer \cite{Meyer1971} showed that between
finite automata and general context-free grammars for regular
languages the difference in economy of description can
be arbitrary, that is, the trade-off is not upper bounded
by any recursive function.
Hartmanis \cite{Hartmanis1980} showed that the descriptional
complexity trade-off between deterministic and nondeterministic
PDA is non-recursive, even if the nondeterministic
PDA is equipped with a proof in a formal system that it defines
a deterministic language. The proof is based on the fundamental
idea due to Hartmanis
that invalid computations of a Turing machine can be encoded
as a context-free language \cite{Hartmanis1967}.

A couple of years later, based
on G\"odel's technique for non-recursive shortening of
proofs of formal systems  by additional axioms,
Hartmanis \cite{Hartmanis1983}
extended the method as a general technique for proving
non-recursive succinctness trade-offs.
 We state the result on non-recursive
 trade-offs using the formulation by Kutrib
 \cite{Kutrib2005} that is independent of a particular
 complexity measure.

\begin{theorem}[\cite{Hartmanis1983,Kutrib2005}]
	\label{tatta1}
	Let $S_1$ and $S_2$ be descriptional systems for
	recursive languages.  The trade-off between $S_1$ and
	$S_2$ is non-recursive if the following conditions hold.

	There exists a
	descriptional system $S_3$
	and a property $P$ that is not semi-decidable
	for languages with a representation in $S_3$ such that,
	given an arbitrary representation ${\cal R} \in S_3$,
	there exists an effective procedure to construct
	a representation in $S_1$ for some language $L_{\cal R}$ with
	the property that
	$L_{\cal R}$ has a representation in $S_2$ if and only
	if  $L({\cal R})$ does not have property $P$.
\end{theorem}

In fact, as noted in \cite{Kutrib2005} most
 proofs appearing in the
 literature to establish non-recursive
 trade-offs rely on a  technique analogous to Theorem~\ref{tatta1}, or
 are based on
  context-free language encodings of invalid Turing machine computations
 from \cite{Hartmanis1967}.

\section{Ambiguity of NFAs and state complexity}

The state complexity ${\rm sc}(L)$ (respectively, nondeterministic state
complexity ${\rm nsc}(L)$) of a regular language $L$ is the minimal number
of states of a DFA
(respectively, of an NFA) for $L$.
Already from \cite{Lupanov1963,Meyer1971,Moore1971} it is known that, for
some regular languages $L$, ${\rm sc}(L) = 2^{{\rm nsc}(L)}$.

\medskip
The degree of ambiguity of an NFA $A$ on a string $w$ is the number
of accepting computations of $A$ on $w$. The NFA $A$ is {\em unambiguous\/}
(UFA) if any string has at most one accepting computation.
If the ambiguity of $A$ on any string is bounded by a constant,
$A$ is {\em finitely
ambiguous\/} (FNFA) and $A$ is {\em polynomially ambiguous\/}
(PNFA) if the
degree of ambiguity of $A$ on input $w$ is bounded by a polynomial in
the length of $w$.

Schmidt \cite{Schmidt1978} developed methods to prove lower bounds
for the size of UFAs and showed that there exists an $n$-state UFA
where the smallest equivalent DFA requires $2^{\Omega(\sqrt{n})}$ states.
The lower bound was improved by different authors and
Leiss \cite{Leiss1981} gives a construction of an $n$-state UFA with
multiple initial states where an equivalent DFA need $2^n$ states.
Leung~\cite{Leung2005} established the lower bound $2^n$ for
determinization of an UFA with one initial state, as well
as, showed that there exists an $n$ state
FNFA for which any equivalent UFA needs $2^n - 1$ states.

Ravikumar and Ibarra \cite{Ravikumar1989} first considered
systematically succinctness comparisons between FNFAs, PNFAs
and general NFAs, and showed that any NFA
recognizing a bounded language can be converted to an FNFA with
polynomial size blow-up. Leung \cite{Leung1998}
gave an optimal  separation between PNFAs and general NFAs.
Using communication complexity Hromkovi\v{c} et al.~\cite{Hromkovic2002}
give a significantly simplified proof for a super-polynomial
separation of NFAs and PNFAs, however, their proof does
not give the exact optimal size blow-up $2^n - 1$.

\begin{theorem}[\cite{Leung1998}]
	For $n \in \mathbb{N}$ there exists an $n$-state  NFA
	$A_n$ such that any PNFA for the language $L(A_n)$ needs
	$2^n - 1$ states.
\end{theorem}

Ravikumar and Ibarra \cite{Ravikumar1989} also conjectured that
polynomially ambiguous NFAs can be significantly more succinct
than finitely ambiguous NFAs. The question was solved
affirmatively by Hrom\-kovi\v{c} and Schnitger \cite{Hromkovic2011}.
 The below theorem gives a
simplified special case of the result in \cite{Hromkovic2011}
that gives a superpolynomial succinctness separation
between NFAs
with degree of ambiguity, respectively,
$O(m^{k-1})$ and $O(m^{k})$, $k \in \mathbb{N}$.
However, the lower bound for the separation of $n$-state
PNFAs and FNFAs is not $2^{\Theta(n)}$ and the precise
trade-off in the economy of description remains open.

\begin{theorem}[\cite{Hromkovic2011}]
	For $n \in \mathbb{N}$ there exists a PNFA $A_n$ with
	number of states polynomial in $n$ such that
	any FNFA recognizing the language $L(A_n)$ has at least
$2^{\Omega(n^\frac{1}{3})}$ states.
\end{theorem}

Besides ambiguity the degree of nondeterminism can be measured,
roughly speaking,
by counting the number of guesses in one computation
\cite{Goldstine1994} or by counting
the number of all computations. The {\em tree width\/}, a.k.a.
{\em leaf size\/} or {\em path size\/}
of an NFA $A$ on input $w$ is the number of leaves
of the computation tree of $A$ on $w$
\cite{Bjorklund2012,Han2017,Hromkovic2002b,Palioudakis2012}.
It is easy to see that an NFA with finite tree width can be
determinized with polynomial size blow-up but
very little is known about succinctness comparisons of NFAs
with different non-constant tree width growth rates.
For example, it remains open whether an NFA with polynomial tree width
may, in the worst case, require super-polynomially more states than
an equivalent unrestricted NFA. Similarly, the succinctness comparison
between NFAs, respectively, of finite and polynomial tree width
remains open.

\section{Operational state complexity}

The effect of a regularity
preserving operation $f$ on the size of the minimal DFA (respectively,
on the size of a minimal NFA) is the {\em operational state complexity\/}
of the operation. This is defined formally below.

\begin{definition}
	\label{datta1}
	If $f$ is an $m$-ary regularity
preserving language operation,
a (deterministic) {\em state complexity upper bound\/} of
$f$ is a function $g : \mathbb{N}^m \rightarrow \mathbb{N}$
such that for any regular languages $L_1$, \ldots, $L_m$, the language
$f(L_1, \ldots, L_m)$ has a DFA with at most
$g({\rm sc}(L_1), \ldots, {\rm sc}(L_m))$ states.
\end{definition}

The nondeterministic state
complexity of an operation $f$ is defined similarly.
A function $f_{\rm sc} : \mathbb{N}^m \rightarrow \mathbb{N}$ is the
precise worst-case state complexity of $f$ if $f_{\rm sc}$ is a state
complexity upper bound of $f$ and, furthermore,
for any positive integers $n_1, \ldots, n_m$ there exist regular
languages $L_i$ with ${\rm sc}(L_i) = n_i$, $i = 1, \ldots, m$,
and the minimal DFA for $f(L_1, \ldots, L_m)$ has
$f_{\rm sc}(n_1, \ldots, n_m)$ states.

The state complexity of language operations was
first considered by Maslov~\cite{Maslov1970} but the paper remained
unknown in the west. A systematic study of operational state
complexity of regular languages was initiated by S. Yu
in the 1990's \cite{Yu1997,Yu2001}. The operational state
complexity of extensions of finite automata that have strong
closure properties, such as input-driven pushdown automata, a.k.a. visibly
pushdown automata, has also  been considered \cite{Alur2004,Okhotin2014}.

In a series of papers
Yu and co-authors have investigated
 the state complexity of combined operations and have determined the precise
 worst-case state complexity of all combinations of two basic
 language operations \cite{Cui2012, Gao2017}. Establishing matching upper
 and lower bounds for the
 state complexity of combined language operations is often involved and
 \'{E}sik et al. \cite{Esik2009} have introduced techniques
 to estimate the state complexity of combined operations.
 For a general combination  of operations
 that include marked concatenation
 and intersection, Yu et al.~\cite{Salomaa2013} have shown that
 the question whether a given integer
 function is a state complexity upper bound
 is  undecidable in the following sense.

\medskip
The marked concatenation of languages $L_1$, $L_2$, \ldots,
$L_n$ is defined as $L_1 \sharp L_2 \sharp \cdots \sharp L_n$
where $\sharp$ is a new symbol not appearing in the languages $L_i$.
A $(\cap, \sharp)$-composition over the set
$\{ L_1, L_2, \ldots, L_n \}$, $n \geq 2$, of language variables
is an expression $\beta_1 \sharp \beta_2 \sharp \cdots \sharp \beta_r$,
$r \geq 2$, where each $\beta_i$ is of the form
$$
\beta_i = K_1 \cap K_2 \cap \cdots \cap K_{t_i}, \;\;  1 \leq t_i \leq n,
$$
where $K_j$'s are distinct among the language variables $L_i$,
$i = 1, \ldots, n$.
A sequence of $(\cap, \sharp)$-compositions
$C_i$, $i = 1, 2, \ldots,$ is effectively constructible if there is
an algorithm that on input $i \in \mathbb{N}$ outputs $C_i$.

\begin{theorem}[\cite{Salomaa2013}]
A sequence of
$(\cap, \sharp)$-compositions $C_i$,
 can be
effectively constructed such that, given $i \in \mathbb{N}$ and
a polynomial with positive integer coefficients $P$ over the
same number of variables as $C_i$,
it is undecidable whether or not $P$ is a state complexity
upper bound for the composition $C_i$
(as defined in Definition~\ref{datta1}).
\end{theorem}


\begin{thebibliography}{10}
\providecommand{\url}[1]{\texttt{#1}}
\providecommand{\urlprefix}{URL }
\expandafter\ifx\csname urlstyle\endcsname\relax
  \providecommand{\doi}[1]{doi:\discretionary{}{}{}#1}\else
  \providecommand{\doi}{doi:\discretionary{}{}{}\begingroup
  \urlstyle{rm}\Url}\fi
\providecommand{\eprint}[2][]{\url{#2}}

\bibitem{Hartmanis1983}
Hartmanis J.
\newblock On {G}{\"o}del speed-up and succinctness of language representations.
\newblock \emph{Theoretical Computer Science}, 1983.
\newblock \textbf{26}(3):335--342.
\newblock \doi{10.1016/0304-3975(83)90016-6}.

\bibitem{Lupanov1963}
Lupanov OB.
\newblock A comparison of two types of finite sources (O sravnenii dvukh tipov
  konechnykh istochnikov).
\newblock \emph{Problemy Kibernetiki}, 1963.
\newblock \textbf{9}:321--326.

\bibitem{Maslov1970}
Maslov AN.
\newblock Estimates of the number of states of finite automata.
\newblock \emph{Soviet Math. Dokl.}, 1970.
\newblock \textbf{11}(5):1373--1375.

\bibitem{Meyer1971}
Meyer AR, Fischer MJ.
\newblock Economy of description by automata, grammars, and formal systems.
\newblock In: Proceedings of {SWAT} 1971. {IEEE}, 1971 pp. 188--191.
\newblock \doi{10.1109/SWAT.1971.11}.

\bibitem{Moore1971}
Moore FR.
\newblock On the bounds for state-set size in the proofs of equivalence between
  deterministic, nondeterministic, and two-way finite automata.
\newblock \emph{{IEEE} Trans. Comput.}, 1971.
\newblock \textbf{{C}-20}(10):1211--1214.
\newblock \doi{10.1109/T-C.1971.223108}.

\bibitem{Stearns1967}
Stearns RE.
\newblock A regularity test for pushdown machines.
\newblock \emph{Information and Control}, 1967.
\newblock \textbf{11}(3):323--340.
\newblock \doi{10.1016/S0019-9958(67)90591-8}.

\bibitem{Sakoda1978}
Sakoda WJ, Sipser M.
\newblock Nondeterminism and the size of two way finite automata.
\newblock In: Proceedings of {STOC} 1978. {ACM}, 1978 pp. 275--286.
\newblock \doi{10.1145/800133.804357}.

\bibitem{Berman1980}
Berman P.
\newblock A note on sweeping automata.
\newblock In: Proceedings of {ICALP} 1980, volume~85 of \emph{{LNCS}}.
  Springer, 1980 pp. 91--97.
\newblock \doi{10.1007/3-540-10003-2_62}.

\bibitem{Sipser1980}
Sipser M.
\newblock Lower bounds on the size of sweeping automata.
\newblock \emph{Journal of Computer and System Sciences}, 1980.
\newblock \textbf{21}(2):195--202.
\newblock \doi{10.1016/0022-0000(80)90034-3}.

\bibitem{Hromkovic2003}
Hromkovi{\v c} J, Schnitger G.
\newblock Nondeterminism versus determinism for two-way finite automata:
  Generalizations of {S}ipser's separation.
\newblock In: Proceedings of {ICALP} 2003, volume 2719 of \emph{{LNCS}}.
  Springer, 2003 pp. 439--451.
\newblock \doi{10.1007/3-540-45061-0_36}.

\bibitem{Kapoutsis2014}
Kapoutsis CA.
\newblock Two-way automata versus logarithmic space.
\newblock \emph{Theory of Computing Systems}, 2014.
\newblock \textbf{55}:421--447.
\newblock \doi{10.1007/s00224-013-9465-0}.

\bibitem{Pighizzini2013}
Pighizzini G.
\newblock Two-way finite automata: Old and recent results.
\newblock \emph{Fundamenta Informaticae}, 2013.
\newblock \textbf{126}(2--3):225--246.
\newblock \doi{10.3233/FI-2013-879}.

\bibitem{Gao2017}
Gao Y, Moreira N, Reis R, Yu S.
\newblock A survey on operational state complexity.
\newblock \emph{Journal of Automata, Languages and Combinatorics}, 2017.
\newblock \textbf{21}(4):251--310.
\newblock \doi{10.25596/jalc-2016-251}.

\bibitem{Goldstine2002}
Goldstine J, Kappes M, Kintala CMR, Leung H, Malcher A, Wotschke D.
\newblock Descriptional complexity of machines with limited resources.
\newblock \emph{Journal of Universal Computer Science}, 2002.
\newblock \textbf{8}(2):193--234.
\newblock \doi{10.3217/jucs-008-02-0193}.

\bibitem{Gruber2015}
Gruber H, Holzer M.
\newblock From finite automata to regular expressions and back --- A summary on
  descriptional complexity.
\newblock \emph{International Journal of Foundations of Computer Science},
  2015.
\newblock \textbf{26}(8):1009--1040.
\newblock \doi{10.1142/S0129054115400110}.

\bibitem{Gruber2021}
Gruber H, Holzer M, Kutrib M.
\newblock Descriptional complexity of regular languages.
\newblock In: Pin J{\'E} (ed.), Handbook of Automata Theory, volume~1, pp.
  411--457. {EMS} Press, 2021.
\newblock \doi{10.4171/AUTOMATA-1/12}.

\bibitem{Holzer2011b}
Holzer M, Kutrib M.
\newblock Descriptional and computational complexity of finite automata---A
  survey.
\newblock \emph{Information and Computation}, 2011.
\newblock \textbf{209}(3):456--470.
\newblock \doi{10.1016/j.ic.2010.11.013}.

\bibitem{Hromkovic2002b}
Hromkovi{\v c} J.
\newblock Descriptional complexity of finite automata: Concepts and open
  problems.
\newblock \emph{Journal of Automata, Languages and Combinatorics}, 2002.
\newblock \textbf{7}(4):519--531.
\newblock \doi{10.25596/jalc-2002-519}.

\bibitem{Yu1997}
Yu S.
\newblock Regular languages.
\newblock In: Rozenberg G, Salomaa A (eds.), Handbook of Formal Languages,
  volume~1, pp. 41--110. Springer, 1997.
\newblock \doi{10.1007/978-3-642-59136-5_2}.

\bibitem{Hartmanis1980}
Hartmanis J.
\newblock On the succinctness of different representations of languages.
\newblock \emph{{SIAM} Journal of Computing}, 1980.
\newblock \textbf{9}(1):114--120.
\newblock \doi{10.1137/0209010}.

\bibitem{Hartmanis1967}
Hartmanis J.
\newblock Context-free languages and {T}uring machine computations.
\newblock In: Mathematical Aspects of Computer Science, volume~19 of
  \emph{Proc. Sympos. Appl. Math.} American Mathematical Society, 1967 pp.
  42--51.
\newblock \doi{10.1090/psapm/019/0235938}.

\bibitem{Kutrib2005}
Kutrib M.
\newblock The phenomenon of non-recursive trade-offs.
\newblock \emph{International Journal of Foundations of Computer Science},
  2005.
\newblock \textbf{16}(5):957--973.
\newblock \doi{10.1142/S0129054105003406}.

\bibitem{Schmidt1978}
Schmidt EM.
\newblock Succinctness of descriptions of context-free, regular, and finite
  languages.
\newblock Ph.D. thesis, Cornell University, 1978.

\bibitem{Leiss1981}
Leiss E.
\newblock Succinct representation of regular languages by boolean automata.
\newblock \emph{Theoretical Computer Science}, 1981.
\newblock \textbf{13}(3):323--330.
\newblock \doi{10.1016/S0304-3975(81)80005-9}.

\bibitem{Leung2005}
Leung H.
\newblock Descriptional complexity of {NFA} of different ambiguity.
\newblock \emph{International Journal of Foundations of Computer Science},
  2005.
\newblock \textbf{16}(5):975--984.
\newblock \doi{10.1142/S0129054105003418}.

\bibitem{Ravikumar1989}
Ravikumar B, Ibarra OH.
\newblock Relating the type of ambiguity of finite automata to the succinctness
  of their representation.
\newblock \emph{{SIAM} Journal of Computing}, 1989.
\newblock \textbf{18}(6):1263--1282.
\newblock \doi{10.1137/0218083}.

\bibitem{Leung1998}
Leung H.
\newblock Separating exponentially ambiguous finite automata from polynomially
  ambiguous finite automata.
\newblock \emph{{SIAM} Journal of Computing}, 1998.
\newblock \textbf{27}(4):1073--1082.
\newblock \doi{10.1137/S0097539793252092}.

\bibitem{Hromkovic2002}
Hromkovi{\v c} J, Seibert S, Karhum{\"a}ki J, Klauck H, Schnitger G.
\newblock Communication complexity method for measuring nondeterminism in
  finite automata.
\newblock \emph{Information and Computation}, 2002.
\newblock \textbf{172}(2):202--217.
\newblock \doi{10.1006/inco.2001.3069}.

\bibitem{Hromkovic2011}
Hromkovi{\v c} J, Schnitger G.
\newblock Ambiguity and communication.
\newblock \emph{Theory of Computing Systems}, 2011.
\newblock \textbf{48}:517--534.
\newblock \doi{10.1007/s00224-010-9277-4}.

\bibitem{Goldstine1994}
Goldstine J, Kintala CMR, Wotschke D.
\newblock On measuring nondeterminism in regular languages.
\newblock \emph{Information and Computation}, 1990.
\newblock \textbf{86}(2):179--194.
\newblock \doi{10.1016/0890-5401(90)90053-K}.

\bibitem{Bjorklund2012}
Bj{\"o}rklund H, Martens W.
\newblock The tractability frontier for {NFA} minimization.
\newblock \emph{Journal of Computer and System Sciences}, 2012.
\newblock \textbf{78}(1):198--210.
\newblock \doi{10.1016/j.jcss.2011.03.001}.

\bibitem{Han2017}
Han YS, Salomaa A, Salomaa K.
\newblock Ambiguity, nondeterminism and state complexity of finite automata.
\newblock \emph{Acta Cybernetica}, 2017.
\newblock \textbf{23}(1):141--157.
\newblock \doi{10.14232/actacyb.23.1.2017.9}.

\bibitem{Palioudakis2012}
Palioudakis A, Salomaa K, Akl SG.
\newblock State complexity of finite tree width {NFA}s.
\newblock \emph{Journal of Automata, Languages and Combinatorics}, 2012.
\newblock \textbf{17}(2--4):245--264.
\newblock \doi{10.25596/jalc-2012-245}.

\bibitem{Yu2001}
Yu S.
\newblock State complexity of regular languages.
\newblock \emph{Journal of Automata, Languages and Combinatorics}, 2001.
\newblock \textbf{6}(2):221--234.
\newblock \doi{10.25596/jalc-2001-221}.

\bibitem{Alur2004}
Alur R, Madhusudan P.
\newblock Visibly pushdown languages.
\newblock In: Proceedings of {STOC} 2004. {ACM}, 2004 pp. 202--211.
\newblock \doi{10.1145/1007352.1007390}.

\bibitem{Okhotin2014}
Okhotin A, Salomaa K.
\newblock Complexity of input-driven pushdown automata.
\newblock \emph{{ACM} {SIGACT} News}, 2014.
\newblock \textbf{45}(2):47--67.
\newblock \doi{10.1145/2636805.2636821}.

\bibitem{Cui2012}
Cui B, Gao Y, Kari L, Yu S.
\newblock State complexity of combined operations with two basic operations.
\newblock \emph{Theoretical Computer Science}, 2012.
\newblock \textbf{437}:82--102.
\newblock \doi{10.1016/j.tcs.2012.02.030}.

\bibitem{Esik2009}
{\'E}sik Z, Gao Y, Liu G, Yu S.
\newblock Estimation of state complexity of combined operations.
\newblock \emph{Theoretical Computer Science}, 2009.
\newblock \textbf{410}(35):3272--3280.
\newblock \doi{10.1016/j.tcs.2009.03.026}.

\bibitem{Salomaa2013}
Salomaa A, Salomaa K, Yu S.
\newblock Undecidability of state complexity.
\newblock \emph{International Journal of Computer Mathematics}, 2013.
\newblock \textbf{90}(6):1310--1320.
\newblock \doi{10.1080/00207160.2012.704994}.
\end{thebibliography}
\end{document}